\documentclass[%
 reprint,
superscriptaddress,
bibnotes,
 amsmath,amssymb,
 aps,
prb,
]{revtex4-2}

\usepackage{graphicx}
\usepackage{dcolumn}
\usepackage{bm}

\begin{document}

\preprint{APS/123-QED}

\title{Dependences of local density of state on temperature, size, and shape in two-dimensional nano-structured superconductors}

\author{Saoto Fukui}
\author{Zhen Wang}%
\affiliation{Shanghai Institute of Microsystem and Information Technology, Chinese Academy of Sciences, Shanghai 200050, China}%


\author{Masaru Kato}
\affiliation{%
 Department of Mathematical Sciences, Osaka Prefecture University, Osaka 599-8531, Japan
}%



\begin{abstract}
In this paper, we investigate a local density of state (LDOS) in two-dimensional nano-structured superconductors.
We solve the Bogoliubov-de Gennes equations self-consistently with the two-dimensional finite element method.
In nano-structured superconductors, the LDOS as a function of the energy has many discrete peaks.
A discretization of the LDOS comes from a discretization of energy levels due to the quantum confinement effect in nano-structured systems.
When the temperature increases, a width of a peak in the LDOS is spread to a large energy range and neighbor peaks are superposed due to the thermal effect.
On the other hand, for the fixed temperature, the behavior of the LDOS is different between  nano-scaled rectangular and square systems.
In the nano-scaled rectangular system, when only a lateral length increases, a contribution of the quantum confinement effect from the lateral side is suppressed, while the contribution from a longitudinal side remains large.
Then, some peaks are left in the LDOS even when the lateral length is very large.
These peaks form a periodic structure and can be regarded as gaps in a multi gap structure due to the quantum confinement effect.
On the other hand, energy levels in the square system tend to arrange equally.
Then, in the square system, peaks in the LDOS which exist in the rectangular system are small.
Also, the period between peaks in the LDOS in the square system is smaller than the period in the rectangular system.
\end{abstract}

\maketitle


\section{\label{introduction} Introduction}
Recently, high quality small superconductors has been developed \cite{savolainen, tian}.
Nanoscopic and mesoscopic superconductors have many interesting phenomena which do not occur in bulk superconductors.
For example, vortices in a small type-II superconductor are confined in a narrow region, which lead to geometry-dependent arrangements of vortices and a formation of a giant vortex state \cite{kokubo,baelus,cren}.
Also, when a thickness of a film is atomic scale or a few nanometer scale, a movement of an electron is restricted strongly.
A confinement of electrons results in a discretization of energy levels.
The discretization of energy levels is one of features of a quantum confinement effect.
In the field of semiconductors, a quantum computer with quantum dots \cite{loss} and a quantum dot semiconductor laser \cite{goulding} are reported as applications of the quantum confinement effect.

The quantum confinement effect is an important phenomenon for not only semiconductors but also superconductors.
It is known that superconductivity is enhanced in nano-particle superconductors \cite{abeles,li_1,li_2,halperin}.
Moreover, basic superconducting properties such as a critical temperature, a density of state (DOS), and a gap energy oscillate when the thickness of the film decreases.
Oscillations of superconducting properties are reported theoretically \cite{blatt,wei,shanenko_2006,shanenko_2007,croitoru,chen,bermudez} and experimentally \cite{guo,ozer,eom}.
As well as vortex arrangements in nano-structured superconductors, the quantum confinement effect depends on the geometry of the sample.
Then, oscillations of superconducting properties depend on not only the system size but also the geometry \cite{croitoru}. 
To investigate relations between geometrical influences and superconducting properties is important for applications of nano-structured superconducting devices such as a nano-SQUID \cite{troeman}, a quantum computer \cite{friesen}, and a single flux quantum (SFQ) logic \cite{yorozu}.
 
The discretization of energy levels in the quantum confinement effect also affects a dependence of a local density of state (LDOS) on the energy dramatically.
In a LDOS-energy plot, a discretization of the LDOS due to the discretization of energy levels is reported in hollow nano-cylinders \cite{chen} and superconducting nano-ribbons with constrictions \cite{flammia}.
Experimentally, the LDOS corresponds to a differential conductance with a STM/STS measurement.
In the STM/STS measurement, the differential conductance is measured by a tunneling current between a probe of a microscopy and a surface in the sample.
Then, an electric structure in the two-dimensional surface is important for the LDOS.
In two-dimensional nano-structured systems, a correlation among the dependence of the LDOS on the energy, the system condition, and the discretization of energy levels is hardly reported. 
In this paper, we show dependences of the LDOS on the temperature, the size, and the shape in two-dimensional superconductors, in particular, rectangular and square systems.
Also, we investigate relations between the LDOS and the discretization of energy levels in detail.

This paper is organized as follows.
In Sec.~\ref{method}, we introduce our model, the Bogoliubov-de Gennes equations, and a numerical method.
In Sec.~\ref{result}, we show the LDOS in the nano-structured finite system.
First, we show the spatial inhomogeneous gap energy and the LDOS.
Next, dependences of the LDOS on the temperature, the size, and the shape in the system are discussed.
We also mention dependences of the discretization of energy levels on the size and the shape.
In Sec.~\ref{summary}, we give a brief summary of this paper.

\section{\label{method} Method}
We consider two-dimensional conventional ($s$-wave) superconductors.
We use the microscopic Bogoliubov-de Gennes (BdG) equations in order to investigate electronic structures \cite{de_gennes}.
The BdG equations are given by,
\begin{subequations}
\begin{eqnarray}
 & & \left[ \frac{1}{2m} \left( \frac{\hbar}{i} \nabla - \frac{e\mbox{\boldmath $A$}}{c} \right)^2 - \mu \right] u_n(\mbox{\boldmath $r$}) + \Delta(\mbox{\boldmath $r$}) v_n(\mbox{\boldmath $r$}) \nonumber \\
 & & = E_n u_n(\mbox{\boldmath $r$}), \label{bdg_1} \\
 & & -\left[ \frac{1}{2m} \left( \frac{\hbar}{i} \nabla + \frac{e\mbox{\boldmath $A$}}{c} \right)^2 - \mu \right] v_n(\mbox{\boldmath $r$}) + \Delta^\ast(\mbox{\boldmath $r$}) u_n(\mbox{\boldmath $r$}) \nonumber \\
 & & = E_n v_n(\mbox{\boldmath $r$}), \label{bdg_2}
\end{eqnarray} \label{bdg}
\end{subequations}
where $u_n(\mbox{\boldmath $r$})$ and $v_n(\mbox{\boldmath $r$})$ are $n$-th quasi-particle wave functions, $E_n$ is a $n$-th eigenenergy measured with respect to a chemical potential, $\Delta(\mbox{\boldmath $r$})$ is a gap energy, $\mu$ is the chemical potential, $m$ is an electronic effective mass, $e$ is an electronic charge, $c$ is a light velocity, and $\mbox{\boldmath $A$}$ is a magnetic vector potential.
$\Delta(\mbox{\boldmath $r$})$ is obtained from a self-consistent equation,
\begin{equation}
 \Delta(\mbox{\boldmath $r$}) = g \sum_{n}^{|E_n| \leq E_c} u_n(\mbox{\boldmath $r$}) v_n^{\ast}(\mbox{\boldmath $r$}) (1-2f(E_n)). \label{delta}
\end{equation}
Here, $g$ is a BCS coupling constant and $f(E_n) = 1/\{\exp[E_n/(k_BT)]+1\}$ is the Fermi distribution function, where $k_B$ is the Boltzmann constant and $T$ is the temperature.
$E_c$ is a cutoff energy, which corresponds to an energy of a Debye frequency in the Bardeen-Cooper-Schrieffer (BCS) theory, $\hbar \omega_D$ ($\omega_{D}$ is the Debye frequency).
The magnetic vector potential $\mbox{\boldmath $A$}(\mbox{\boldmath $r$})$ is obtained from the Maxwell equations, which is given by,
\begin{equation}
 \nabla \times \left( \nabla \times \mbox{\boldmath $A$}(\mbox{\boldmath $r$}) - \mbox{\boldmath $H$} \right) = \frac{4\pi}{c} \mbox{\boldmath $j$}(\mbox{\boldmath $r$}). \label{maxwell}
\end{equation}
$\mbox{\boldmath $H$}$ is an uniform applied magnetic field.
$\mbox{\boldmath $j$}(\mbox{\boldmath $r$})$ is a current density, which represents \cite{gygi},
\begin{eqnarray}
 \mbox{\boldmath $j$}(\mbox{\boldmath $r$}) &=& \frac{e\hbar}{2mi} \sum_n \left[ f(E_n) u_n^\ast(\mbox{\boldmath $r$}) \left( \nabla - \frac{ie}{\hbar c} \mbox{\boldmath $A$}(\mbox{\boldmath $r$}) \right) u_n(\mbox{\boldmath $r$}) \right. \nonumber \\ 
 & & \left. + (1-f(E_n)) v_n(\mbox{\boldmath $r$}) \left( \nabla - \frac{ie}{\hbar c} \mbox{\boldmath $A$} \right) v_n^\ast(\mbox{\boldmath $r$}) - {\rm H.~C.} \right]. \nonumber \\ \label{current}
\end{eqnarray}
Here, "H.~C." represents the Hermitian conjugate.
In this approach, a particle number must be fixed by controlling the chemical potential $\mu$.
The chemical potential $\mu$ is determined by a particle number conservation,
\begin{equation}
 N_e = 2\int \sum_{n} \left[ |u_n(\mbox{\boldmath $r$})|^2 f(E_n) + |v_n(\mbox{\boldmath $r$})|^2 (1-f(E_n)) \right], \label{particle_number}
\end{equation}
where $N_e$ is a total particle number.

Boundary conditions are that (i) quasi-particle wave functions $u_n(\mbox{\boldmath $r$})$ and $v_n(\mbox{\boldmath $r$})$ are absent at edges of the system, (ii) the current does not flow perpendicular to boundaries, which is expressed by,
\begin{subequations}
\begin{eqnarray}
 \mbox{\boldmath $n$} \cdot \left( \frac{\hbar}{i} \nabla - \frac{e\mbox{\boldmath $A$}}{c} \right) u_n(\mbox{\boldmath $r$}) &=& 0,  \label{boundary_1} \\
 \mbox{\boldmath $n$} \cdot \left( \frac{\hbar}{i} \nabla + \frac{e\mbox{\boldmath $A$}}{c} \right) v_n(\mbox{\boldmath $r$}) &=& 0, \label{boundary_2}
\end{eqnarray}
\end{subequations}
where $\mbox{\boldmath $n$}$ is an unit vector perpendicular to boundaries. 

\begin{figure}[tb]
\begin{tabular}{cc}
\begin{minipage}{0.5\hsize}
\begin{center}
\includegraphics[width=\hsize]{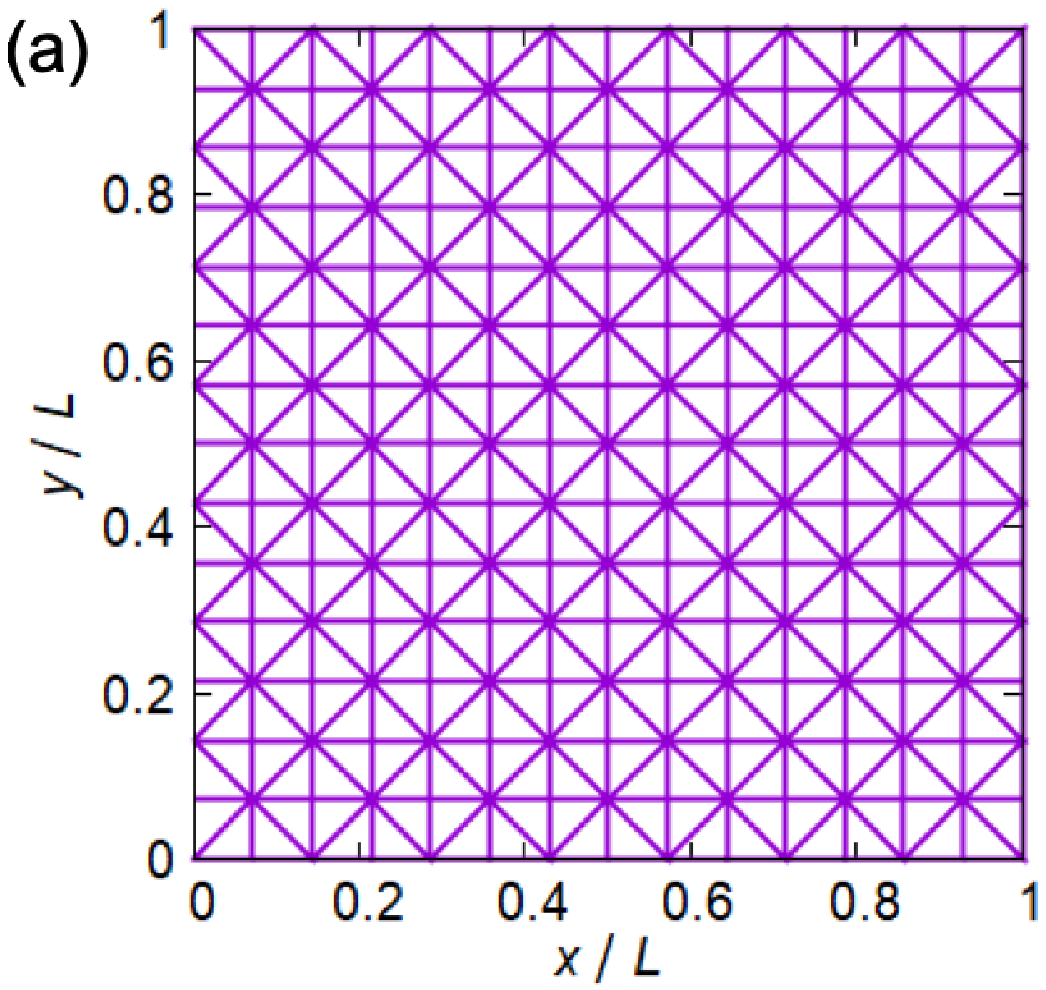}
\label{system_finite_element_a}
\end{center}
\end{minipage}
\begin{minipage}{0.45\hsize}
\begin{center}
\includegraphics[width=\hsize]{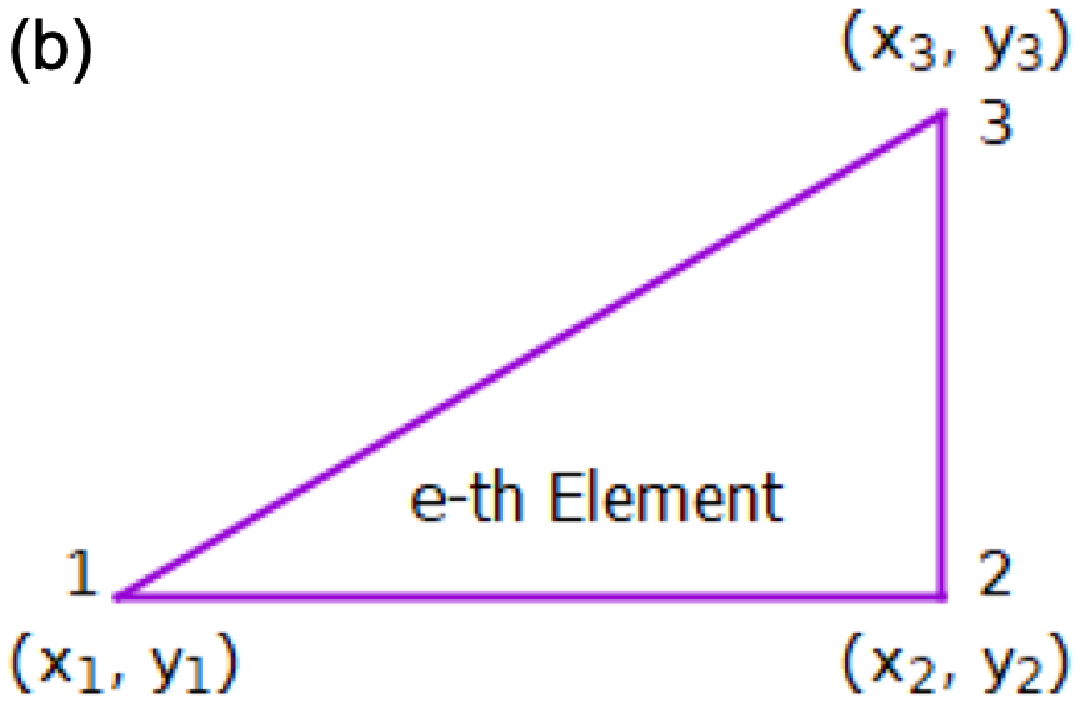}
\label{system_finite_element_b}
\end{center}
\end{minipage}
\end{tabular}
\caption{\label{system_finite_element} (a) Square system divided into triangular elements, (b) Triangular element in the $e$-th element. Coordinates at each node are defined as $1~(x_1,y_1)$, $2~(x_2,y_2)$, and $3~(x_3,y_3)$.}
\end{figure}

In order to solve above equations, we use a finite element method (FEM) \cite{suematsu,kato,umeda, hu}.
In the two-dimensional FEM, we divide a system into small triangular elements, which is shown in Fig. \ref{system_finite_element}(a).
A triangle in Fig.~\ref{system_finite_element}(b) shows one of elements in the system in Fig.~\ref{system_finite_element}(a). 
In the $e$-th triangular element, we label nodes as $1$, $2$, and $3$.
These coordinates in nodes are defined as $\mbox{\boldmath $r$}_1 = (x_1,~y_1)$, $\mbox{\boldmath $r$}_2 = (x_2,~y_2)$, and $\mbox{\boldmath $r$}_3 = (x_3,~y_3)$, respectively.
Then, we define an $e$-th area coordinate at $i$-th node as $\zeta_i^e(x_i,y_i)$, which is given by,
\begin{equation}
 \zeta_i^e(x_i,y_i) = \frac{1}{2S_e}(a_i + b_ix + c_iy), ~~~(i=1,2,3) \label{area_coordinate}
\end{equation}
where $S_e$ is an area of the $e$-th triangular element and,
\begin{subequations}
\begin{eqnarray}
 a_i &=& x_j y_k - x_k y_j, \label{area_coordinate_a} \\
 b_i &=& y_j - y_k, \label{area_coordinate_b} \\
 c_i &=& x_k - x_j. \label{area_coordinate_c}
\end{eqnarray}
\end{subequations}
$i,~j,$ and $k$ are $1,~2,$ and $3$, following a cyclic order.
Using the area coordinate $\zeta_i^e(\mbox{\boldmath $r$})$, we extend quasiparticle wave functions $u_n(\mbox{\boldmath $r$})$ and $v_n(\mbox{\boldmath $r$})$, the gap energy $\Delta(\mbox{\boldmath $r$})$, and the magnetic vector potential $\mbox{\boldmath $A$}(\mbox{\boldmath $r$})$ into,
\begin{subequations}
\begin{eqnarray}
 u_n(\mbox{\boldmath $r$}) &=& \sum_e \sum_i \zeta_i^e(x_i,y_i) u_{n,i}^e, \label{finite_element_u} \\
 v_n(\mbox{\boldmath $r$}) &=& \sum_e \sum_i \zeta_i^e(x_i,y_i) v_{n,i}^e, \label{finite_element_v} \\
 \Delta(\mbox{\boldmath $r$}) &=& \sum_e \sum_i \zeta_i^e(x_i,y_i) \Delta_i^e, \label{finite_element_delta} \\
 \mbox{\boldmath $A$}(\mbox{\boldmath $r$}) &=& \sum_e \sum_i \zeta_i^e(x_i,y_i) \mbox{\boldmath $A$}_i^e. \label{finite_element_a} 
\end{eqnarray} \label{finite_element}
\end{subequations}
We substitute Eqs.~(\ref{finite_element_u})-(\ref{finite_element_a}) into Eqs.~(\ref{bdg})-(\ref{particle_number}).
The BdG equations [Eqs.~(\ref{bdg_1})~and~(\ref{bdg_2})] can be solved as an eigenvalue problem,
\begin{subequations}
\begin{eqnarray}
 & &\sum_j \left[ P_{ij}^{1e}(\mbox{\boldmath $A$}) + P_{ij}^{2e}(\mbox{\boldmath $A$}) \right] u_{n,j}^e + \sum_j Q_{ij}^e(\Delta) v_{n,j}^e \nonumber \\
 & & = E_n \sum_j I_{ij}^e u_{n,j}^e, \label{bdg_fem_1} \\
 & & \sum_j \left[ -P_{ij}^{1e}(\mbox{\boldmath $A$}) + P_{ij}^{2e}(\mbox{\boldmath $A$}) \right] v_{n,j}^e + \sum_j Q_{ij}^{e\ast}(\Delta) u_{n,j}^e  \nonumber \\
 & &= E_n \sum_j I_{ij}^e v_{n,j}^e. \label{bdg_fem_2}
\end{eqnarray} \label{bdg_fem}
\end{subequations}
The Maxwell equation and the current [Eqs.~(\ref{maxwell}) and (\ref{current})] are combined, which is given by,
\begin{subequations}
\begin{eqnarray}
 \sum_j R_{ij}^e(u,v) A_{jx}^e + \sum_{j} S_{ij}^e A_{jy}^e &=& T_i^{ex}(u,v) - U_i^{ey}, \nonumber \\ \label{maxwell_fem_1} \\
 \sum_j R_{ij}^e(u,v) A_{jy}^e - \sum_j S_{ij}^e A_{jx}^e &=& T_i^{ey}(u,v) + U_i^{ex}, \nonumber \\ \label{maxwell_fem_2}
\end{eqnarray} \label{maxwell_fem}
\end{subequations}
where $A_{jx}^e$ and $A_{jy}^e$ are $x$- and $y$-components of the magnetic vector potential $\mbox{\boldmath $A$}_j^e$.
When we derive Eqs.~(\ref{maxwell_fem_1}) and (\ref{maxwell_fem_2}), we impose the London gauge $\nabla \cdot \mbox{\boldmath $A$} = 0$.
Finally, the self-consistent equation [Eq.~(\ref{delta})] and the particle number conservation [Eq.~(\ref{particle_number})] are reproduced by,
\begin{equation}
  \sum_j I_{ij}^e \Delta_j^e = g \sum_{i_1,i_2} I_{ii_1i_2}^e \sum_{n}^{|E_n| \leq E_c} u_{n,i_1}^e v_{n,i_2}^{e\ast} (1-2f(E_n)), \label{delta_fem}
\end{equation}
\begin{eqnarray}
  N_e &=& 2\sum_{ije} I_{ij}^e \sum_n \left[ f(E_n) u_{n,i}^{e\ast} u_{n,j}^e + (1-f(E_n)) v_{n,i}^{e\ast} v_{n,j}^e \right]. \nonumber \\ \label{particle_number_fem}
\end{eqnarray}
Here, we define integrals $I_{ij}^e$, $I_{i_1i_2i_3}^e$, $I_{i_1i_2i_3i_4}^e$, $J_{j}^{ex_i}$, $J_{i_1i_2i_3}^{ex_i}$, and $K_{i_1i_2}^{ex_ix_j}$ as following forms,
\begin{subequations}
\begin{eqnarray}
 I_{ij}^e &\equiv& \int \zeta_i^e(\mbox{\boldmath $r$}) \zeta_j^e(\mbox{\boldmath $r$}) d\mbox{\boldmath $r$}, \label{i2} \\
 I_{i_1i_2i_3}^e &\equiv& \int \zeta_{i_1}^e(\mbox{\boldmath $r$}) \zeta_{i_2}^e(\mbox{\boldmath $r$}) \zeta_{i_3}^e(\mbox{\boldmath $r$}) d\mbox{\boldmath $r$}, \label{i3} \\
 I_{i_1i_2i_3i_4}^e &\equiv& \int \zeta_{i_1}^e(\mbox{\boldmath $r$}) \zeta_{i_2}^e(\mbox{\boldmath $r$}) \zeta_{i_3}^e(\mbox{\boldmath $r$}) \zeta_{i_4}^e(\mbox{\boldmath $r$}) d\mbox{\boldmath $r$}, \label{i4} \\
 J_{j}^{ex_i} &\equiv& \int \frac{\partial \zeta_j^e(\mbox{\boldmath $r$})}{\partial x_i} d\mbox{\boldmath $r$}, \label{j} \\
 J_{i_1i_2i_3}^{ex_i} &\equiv& \int \frac{\partial \zeta_{i_1}^e(\mbox{\boldmath $r$})}{\partial x_i} \zeta_{i_2}^e(\mbox{\boldmath $r$}) \zeta_{i_3}^e(\mbox{\boldmath $r$}) d\mbox{\boldmath $r$}, \label{j3} \\
 K_{i_1i_2}^{ex_i x_j} &\equiv& \int \frac{\partial \zeta_{i_1}^e(\mbox{\boldmath $r$})}{\partial x_i} \frac{\partial \zeta_{i_2}^e(\mbox{\boldmath $r$})}{\partial x_j} d\mbox{\boldmath $r$}, \label{k}
\end{eqnarray}
\end{subequations}
where $x_i = x,~y$.
Using Eqs.~(\ref{i2})-(\ref{k}), we also define coefficients $P_{ij}^{1e}(\mbox{\boldmath $A$})$, $P_{ij}^{2e}(\mbox{\boldmath $A$})$, $Q_{ij}^e(\Delta)$, $R_{ij}^e(u,v)$, $S_{ij}^e$, $T_i^{e\alpha}(u,v)$, and $U_i^{e\alpha}$ as,
\begin{subequations}
\begin{eqnarray}
 P_{ij}^{1e}(\mbox{\boldmath $A$}) &\equiv& \frac{\hbar^2}{2m} \sum_{\alpha} K_{jj}^{e\alpha \alpha} + \frac{e^2}{2mc^2} \sum_{i_1i_2} \sum_{\alpha} I_{iji_1i_2}^e A_{i_1 \alpha}^e A_{i_2 \alpha}^e \nonumber \\ & &  - \mu I_{ij}^e, \label{p1} \\ 
 P_{ij}^{2e}(\mbox{\boldmath $A$}) &\equiv& \frac{ie\hbar}{2mc} \sum_{i_1} \sum_{\alpha} (J_{ii_1j}^{e\alpha} - J_{ji_1i}^{e\alpha}) A_{i_1\alpha}^e, \label{p2} \\
 Q_{ij}^e(\Delta) &\equiv& \sum_{i_1} \Delta_{i_1}^e I_{iji_1}^e, \label{q} \\
 R_{ij}^e(u,v) &\equiv& \sum_{\alpha} K_{ij}^{e\alpha\alpha} + \frac{4\pi e^2}{mc^2} \sum_{i_1i_2} I_{iji_1i_2}^e \nonumber \\ & & \times \sum_n \left[ f(E_n) u_{n,i_1}^{e\ast} u_{n,i_2}^e  \right. \nonumber \\ & & \left.  + (1-f(E_n)) v_{n,i_1}^e v_{n,i_2}^{e\ast} \right], \label{r} \\
 S_{ij}^e &\equiv& K_{ij}^{exy} - K_{ij}^{eyx}, \label{s} \\
 T_i^{e\alpha}(u,v) &\equiv& i \frac{4\pi e\hbar}{2mc} \sum_{i_1i_2} (J_{i_1ii_2}^{e\alpha} - J_{i_2ii_1}^{e\alpha}) \sum_n \left[ f(E_n) u_{n,i_1}^{e\ast} u_{n,i_2}^e \right. \nonumber \\
 & & \left. + (1-f(E_n)) v_{n,i_1}^e v_{n,i_2}^{e\ast} \right], \label{t} \\
 U_i^{e\alpha} &\equiv& H_0 J_i^{e\alpha}, \label{u}
\end{eqnarray} 
\end{subequations}
where $\alpha = x,~y$ and $H_0$ is a magnitude of the applied magnetic field perpendicular to the system [$\mbox{\boldmath $H$} = (0,0,H_0)$].

We solve Eqs.~(\ref{bdg_fem})-(\ref{particle_number_fem}) self-consistently.
First of all, we determine the BCS coupling constant $g$ and the total particle number $N_e$.
We solve Eqs.~(\ref{bdg_fem_1}) and (\ref{bdg_fem_2}) by using a bulk value of the gap energy at the zero temperature ($\Delta_0$) and the chemical potential at $T=0$ ($\mu = \hbar^2k_F^2/2m$, where $k_F$ is a Fermi wave vector).
Substituting solutions in Eqs.~(\ref{bdg_fem_1}) and (\ref{bdg_fem_2}) into Eqs.~(\ref{delta_fem}) and (\ref{particle_number_fem}), $g$ and $N_e$ are obtained.
The coupling coefficient $g$ solved from Eqs.~(\ref{delta_fem}) changes spatially due to the finite system and boundary conditions, which is represented by $g(\mbox{\boldmath $r$})$.
We calculate a spatial average of $g(\mbox{\boldmath $r$})$ by,
\begin{equation}
 g = \frac{1}{S_s} \int_{S_s}g(\mbox{\boldmath $r$}) d\mbox{\boldmath $r$}, \label{g_space}
\end{equation}
where $S_s$ is a superconducting area in the system, in other words, a region except for boundaries in the system.
Next, we set the gap energy $\Delta(\mbox{\boldmath $r$})$ randomly as an initial condition and solve Eqs.~(\ref{bdg_fem_1}) and (\ref{bdg_fem_2}).
After the eigenvalue $E_n$ and corresponding eigenvectors $u_n(\mbox{\boldmath $r$})$ and $v_n(\mbox{\boldmath $r$})$ are obtained, we substitute them to Eqs.~(\ref{maxwell_fem}) - (\ref{particle_number_fem}) in order to obtain $\mbox{\boldmath $A$}(\mbox{\boldmath $r$})$, $\Delta(\mbox{\boldmath $r$})$, and $\mu$.
Then, we substitute solutions in Eqs.~(\ref{maxwell_fem}) - (\ref{particle_number_fem}) into Eqs.~(\ref{bdg_fem_1}) and (\ref{bdg_fem_2}), and obtain the eigenvalue $E_n$ and eigenvectors $u_n(\mbox{\boldmath $r$})$ and $v_n(\mbox{\boldmath $r$})$ again.
This process is iterated until convergence solutions are obtained.

After the self-consistent calculation is finished, we calculate the LDOS $N(\mbox{\boldmath $r$},~E)$.
The LDOS $N(\mbox{\boldmath $r$},~E)$ is given by,
\begin{eqnarray}
 N(\mbox{\boldmath $r$},E) &=& -\sum_n \left[ |u_n(\mbox{\boldmath $r$})|^2 f'(E_n - E) \right. \nonumber \\ &  & \left. + |v_n(\mbox{\boldmath $r$})|^2 f'(E_n + E) \right]. \label{ldos}
\end{eqnarray}
$f'(E) = \partial f(E)/ \partial E$ is a derivative of the Fermi distribution function \cite{gygi, suematsu}. 

\section{\label{result} Results and Discussions}
In this section, we show the LDOS in two-dimensional nano-structured superconductors.
We investigate behaviors of the LDOS in various system sizes.
At first, we set the square system size $L \times L$ in Fig.~\ref{system_finite_element}(a) as a standard system, where $L$ is a length of the standard system.
Then, we consider various system size, for example, $3L \times 3L,~3L\times L,~5L \times L$, and so on.

We set the gap energy in the bulk sample at the zero temperature $\Delta_0 = 0.2 E_c$, the coherence length at the zero temperature $\xi_0 = 0.2 L$, and the Fermi wave vector $k_F = 3.0/\xi_0$, respectively.
Parameters are normalized to the cutoff energy $E_c$ and the length of the system $L$.
Also, we set a Ginzburg-Landau parameter in the clean limit $\kappa = 0.96 \lambda_L/\xi_0 = 3.0$ \cite{de_gennes}, where $\lambda_L$ is a London penetration length.
In this research, the magnetic field is not applied ($H_0 = 0.0$).
In addition, the LDOS is normalized to a density of state (DOS) in the normal state in the two-dimensional system per area $L \times L$ at the zero temperature, $N(0) = m/(\pi^2\hbar^2)$.

\subsection{ \label{op_ldos} Distributions of LDOS in the finite small system}
\begin{figure}[htb]
\centering
\includegraphics[scale=0.5]{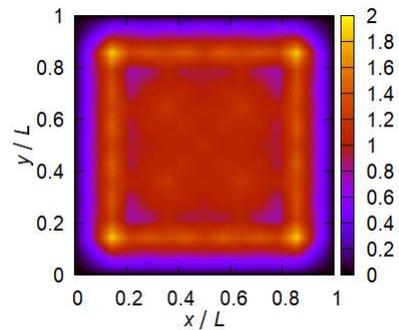}
\caption{\label{op} Spatial distribution of the gap energy $\Delta(\mbox{\boldmath $r$})/\Delta_0$ at $T=0.2T_c$ in the square system $L \times L$ [Fig.~\ref{system_finite_element}(a)].}
\end{figure}

The LDOS relates to wave functions of quasiparticle and the gap energy in Eq.~(\ref{ldos}).
It is known that wave functions and the gap energy are inhomogeneous spatially in finite systems \cite{umeda}.
So, we present a distribution of the gap energy in the finite system before we show results of the LDOS.
We consider the square system, whose size is $L \times L$ [see Fig.~\ref{system_finite_element}(a)] and set the temperature $T = 0.2T_c$, where $T_c$ is the critical temperature in the bulk superconductor.
Fig.~\ref{op} shows the distribution of the gap energy, which is normalized to the gap energy in the bulk superconductor at the zero temperature ($\Delta(\mbox{\boldmath $r$})/\Delta_0$).
Superconducting electrons cannot exist at edges due to boundary conditions, then magnitudes of the gap energy at edges are zero.
Magnitudes of the gap energy are large at locations near boundaries, in particular, at corners $(x/L,~y/L) \sim (0.15,~0.15),~(0.85,~0.15),~(0.15,~0.85),$ and $(0.85,~0.85)$.
\begin{figure}[tb]
\begin{center}
\includegraphics[scale=0.5]{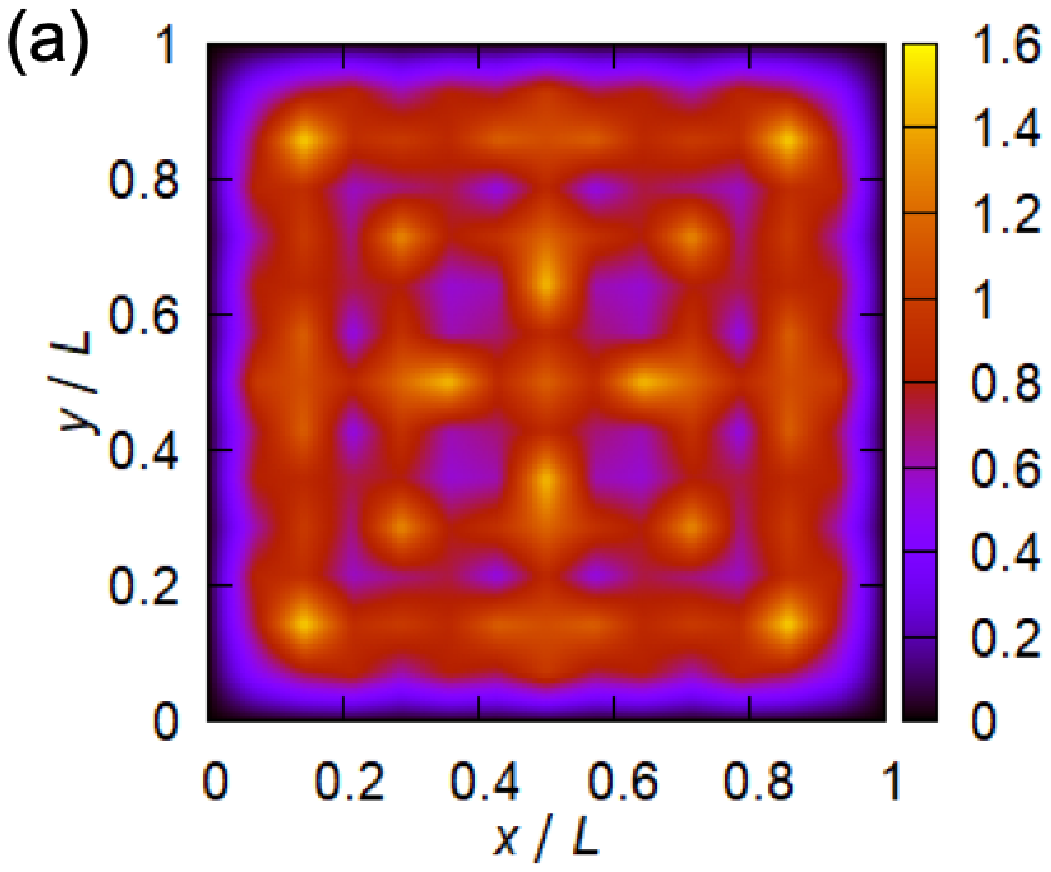}
\label{ldos_space_a}
\end{center}
\begin{center}
\includegraphics[scale=0.5]{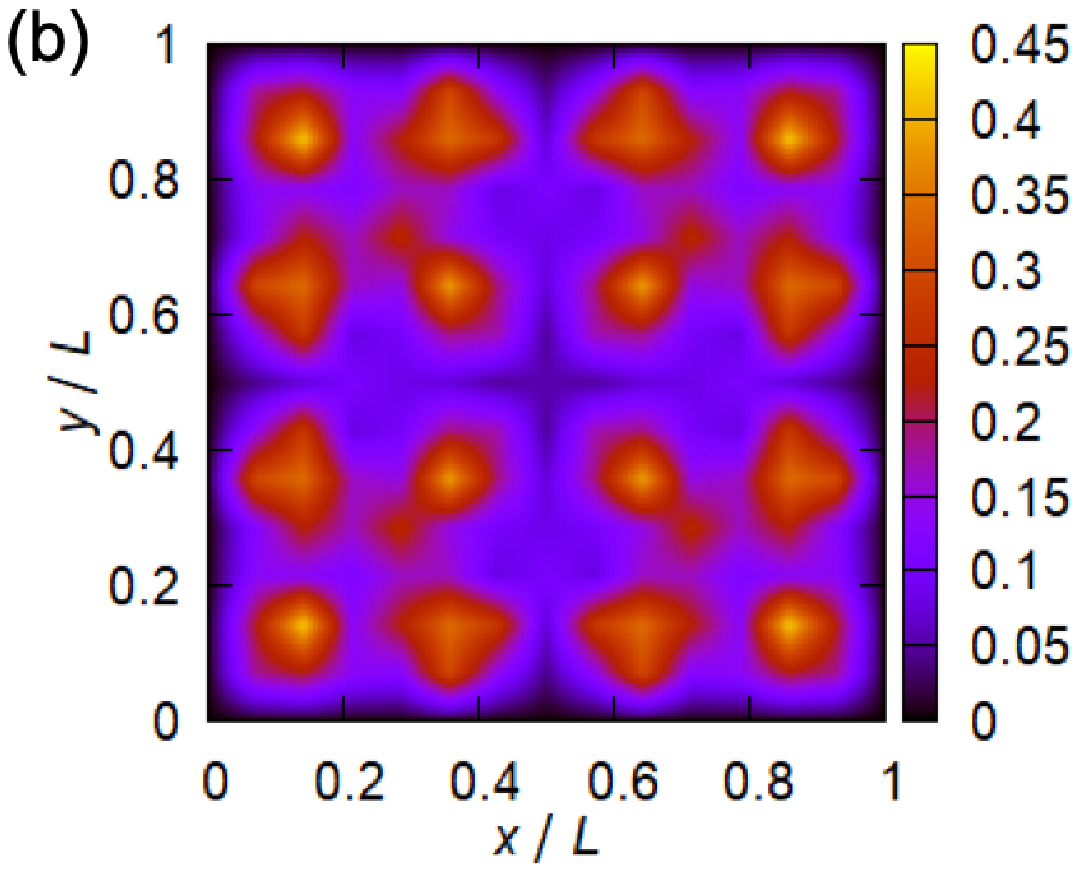}
\label{ldos_space_b}
\end{center}
\caption{\label{ldos_space} Spatial distributions of the LDOS for (a) $E/\Delta_0 = 2.0$ and (b) $E/\Delta_0 = 3.0$ in the square system $L \times L$. The LDOS is normalized to the DOS in the normal state per unit, $N(\mbox{\boldmath $r$}, E)/N(0)$}
\end{figure}
\begin{figure}[tb]
\centering
\includegraphics[scale=0.5]{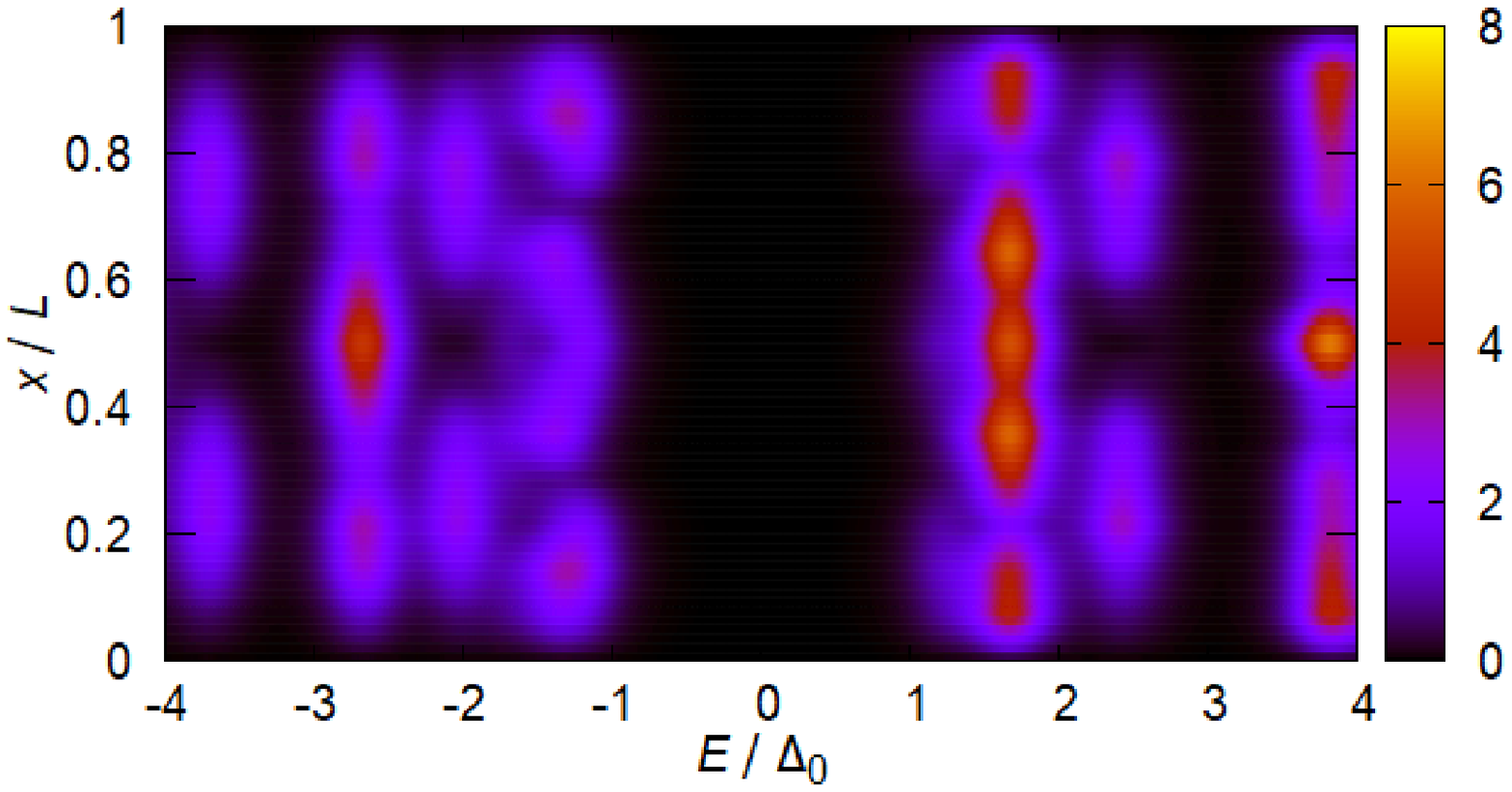}
\caption{\label{ldos_energy} Distribution of the LDOS at middle of the lateral side [$y=0.5L$] in the square system $L \times L$. The LDOS is normalized to the DOS in the normal state per unit, $N(x,y=0.5L,E)/N(0)$}
\end{figure}
\begin{figure}[tb]
\centering
\includegraphics[scale=0.5]{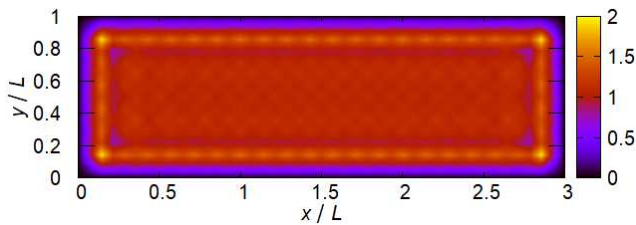}
\caption{\label{op_rec} Spatial distribution of the gap energy $\Delta(\mbox{\boldmath $r$})/\Delta_0$ at $T=0.2T_c$ in the rectangular system $3L \times L$.}
\end{figure}
\begin{figure}[tb]
\begin{center}
\includegraphics[scale=0.5]{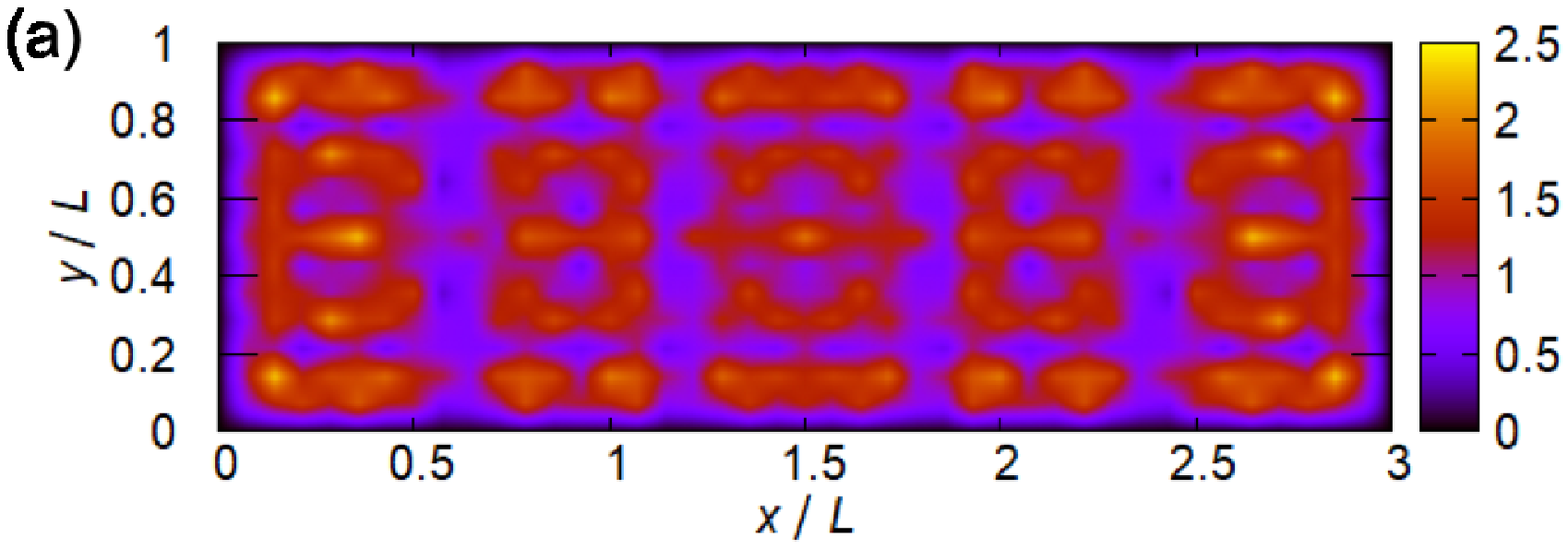}
\label{ldos_space_a}
\end{center}
\begin{center}
\includegraphics[scale=0.5]{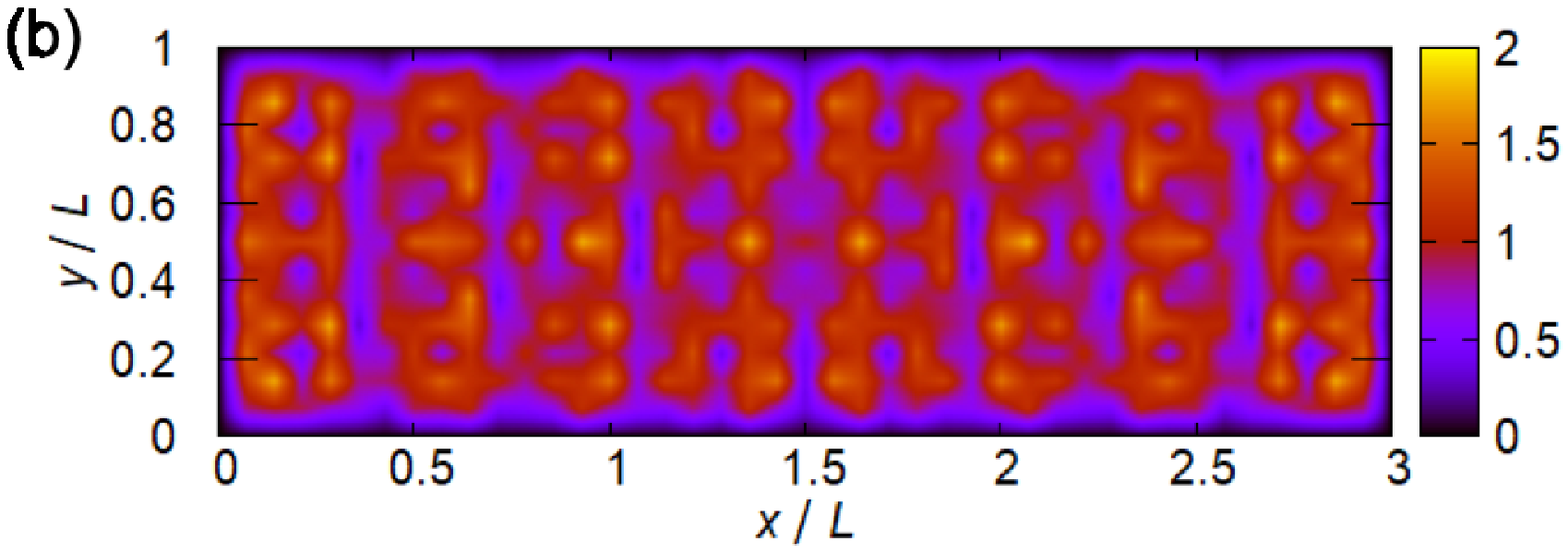}
\label{ldos_space_b}
\end{center}
\caption{\label{ldos_space_rec} Spatial distributions of the LDOS for (a) $E/\Delta_0 = 2.0$ and (b) $E/\Delta_0 = 3.0$ in the rectangular system $3L \times L$. The LDOS is normalized to the DOS in the normal state per unit, $N(\mbox{\boldmath $r$}, E)/N(0)$}
\end{figure}
\begin{figure}[t]
\centering
\includegraphics[scale=0.5]{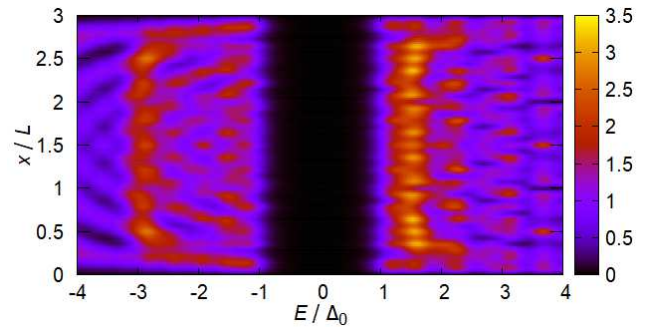}
\caption{\label{ldos_energy_rec} Distribution of the LDOS at middle of the lateral side [$y=0.5L$] in the rectangular system $3L \times L$. The LDOS is normalized to the DOS in the normal state per unit, $N(x,y=0.5L,E)/N(0)$}
\end{figure}

The LDOS has also an inhomogeneous spatial distribution due to the inhomogeneous spatial distribution of the gap energy in Fig.~\ref{op}.
Fig.~\ref{ldos_space} shows spatial distributions of the LDOS for (a) $E/\Delta_0 = 2.0$ and (b) $E/\Delta_0 = 3.0$.
In Fig.~\ref{ldos_space}(a), the LDOS for $E/\Delta_0 = 2.0$ is large at corners, diagonal directions, and cross lines of the middle in the system.
On the other hand, the LDOS is small at corners of the small square region, $(x/L,~y/L) \sim (0.4,~0.4),~(0.6,~0.4),~(0.4,~0.6),$ and $(0.6,~0.6)$.
For the different energy $E/\Delta_0 = 3.0$ in Fig.~\ref{ldos_space}(b), the magnitude of the LDOS is smaller than the magnitude for $E/\Delta_0 = 2.0$ in Fig.~\ref{ldos_space}(a), but locations of peaks are different.
In Fig.~\ref{ldos_space}(b), the LDOS for $E/\Delta_0 = 3.0$ at the cross lines of the middle [$x/L \sim 0.5$ or $y/L \sim 0.5$], where the magnitude of the LDOS is large for $E/\Delta_0 = 2.0$ in Fig.~\ref{ldos_space}(a), is very small.
On the other hand, the LDOS at corners of the small square region $(x/L,~y/L) \sim (0.4,~0.4),~(0.6,~0.4),~(0.4,~0.6),$ and $(0.6,~0.6)$ are almost same as the LDOS for $E/\Delta_0 = 2.0$ in Fig.~\ref{ldos_space}(a) [$N(\mbox{\boldmath $r$},E/\Delta_0=3.0) \sim N(\mbox{\boldmath $r$},E/\Delta_0=2.0) \sim 0.4N(0)$].
Both distributions of the LDOS in Figs~\ref{ldos_space}(a) and (b) are inhomogeneous spatially, but symmetric with respect to middles in systems.

Generally, the DOS in the bulk superconductor has a peak corresponding to the homogeneous gap energy in the positive energy region.
Also, when the magnitude of the energy increases, the DOS decreases continuously as a function of the minus square root \cite{de_gennes, tinkham}. 
On the other hand, in the nano-structured finite system, the behavior of the LDOS as a function of the energy depends on the coordinate in the system. 
This dependence results from the inhomogeneous spatial distribution of the gap energy and the fact that we consider the finite system.
Next, the dependence of the LDOS on the energy is investigated in detail.

Fig.~\ref{ldos_energy} shows the spatial and energy distribution of the LDOS.
The $y$ component of the coordinate is fixed to $y/L =0.5$.
From Fig.~\ref{ldos_energy}, we find some discrete peaks in the LDOS.
The discretization of the LDOS in the vertical direction represents spatial distributions in Fig.~\ref{ldos_space}.
On the other hand, the discretization of the LDOS with respect to the energy results from the different reason.
In nano-structured superconductors, energy levels are discrete due to the quantum confinement effect.
We show a distribution of energy levels in this system in Fig.~\ref{energy_level}, and will discuss it in detail later.
The discretization of energy levels leads to the discretization of the LDOS in Fig.~\ref{ldos_energy}.

In a later subsection (Sec.~\ref{ldos_size_sec}), we will discuss the relation between the LDOS and the system shape, and consider the LDOS in rectangular systems.
In order to compare with the case in the square system, we also show the spatial distribution of the gap energy, the spatial distribution of the LDOS for certain energies, and the spatial and energy distribution of the LDOS in the rectangular system in Figs.~\ref{op_rec},~\ref{ldos_space_rec}, and \ref{ldos_energy_rec}, where their system sizes are fixed to $3L \times L$ and $T = 0.2T_c$.
In Fig ~\ref{op_rec}, as well as the case in the square system, the inhomogeneous spatial distribution of the gap energy is obtained, and magnitudes of the gap energy are large at locations near boundaries.
Figs.~\ref{ldos_space_rec}(a) and (b) show  spatial distributions of the LDOS for $E/\Delta_0 = 2.0$ and $3.0$, respectively, as well as Figs.~\ref{ldos_space}(a) and (b).
Both distributions in Figs.~\ref{ldos_space_rec}(a) and (b) seem to be more inhomogeneous spatially than distributions in the square system in Figs.~\ref{ldos_space}(a) and (b), while these distributions in the rectangular system remain to be symmetric with respect to middles in systems.
Also, as well as the distribution in Fig.~\ref{ldos_energy}, the spatial and energy distribution of the LDOS in Fig.~\ref{ldos_energy_rec} has discrete peaks.
Peaks of the LDOS in the square system in Fig.~\ref{ldos_energy} are large, and their peaks appear very locally. 
On the other hand, the LDOS in the rectangular system in Fig.~\ref{ldos_energy_rec} is smaller than the case in the square system, but more peaks appear.
From this result, a suppression of the discreteness of the LDOS due to the increase of the system size is expected.
The dependence of the discreteness of the LDOS on the system size will be discussed in Sec.~\ref{ldos_size_sec} in detail.

In this subsection, we found inhomogeneous spatial distributions of the LDOS because of inhomogeneous spatial distributions of the gap energy.
Also, the discretization of the LDOS with respect to the energy comes from the discretization of energy levels.
Next, we investigate influences of the temperature, the size, and the shape on the discretization of the LDOS.
In the next subsection, we focus on effects of the temperature on the LDOS.

\subsection{Effects of temperature on the LDOS}
\begin{figure}[tb]
\centering
\includegraphics[scale=0.49]{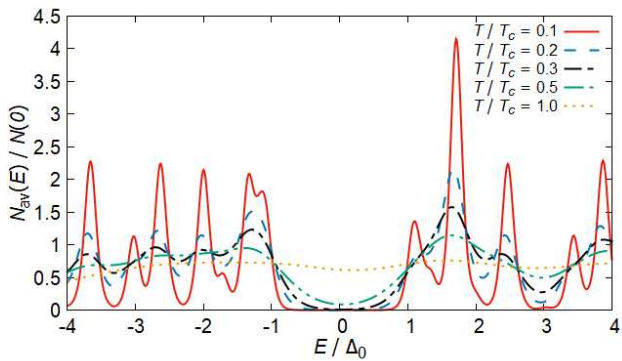}
\caption{\label{ldos_temperature} Space-averaged LDOS as a function of the energy at $T/T_c=0.1$ (red),$~0.2$ (blue),$~0.3$ (black),$~0.5$ (green) and $1.0$ (yellow) in $|E/\Delta_0| < 4.0$. System sizes are fixed to $L \times L$,}
\end{figure}
In this subsection, we investigate dependences of the LDOS on the temperature.
We calculate a spatial-averaged LDOS $N_{\rm av}(E)$, which is given by,
\begin{equation}
 N_{\rm av}(E) = \frac{1}{S} \int_S N(\mbox{\boldmath $r$},~E) d\mbox{\boldmath $r$}, \label{ldos_average}
\end{equation}
where $S$ is a total area in the system, which includes boundaries.
We substitute Eqs.~(\ref{finite_element_u}), (\ref{finite_element_v}), and (\ref{ldos}) into Eq.~(\ref{ldos_average}), then the spatial-averaged LDOS can be rewritten by,
\begin{eqnarray}
 N_{\rm av}(E) &=& -\frac{1}{S} \sum_e \sum_n \sum_{i_1i_2} I_{i_1 i_2}^e  \left\{ u_{n,i_1}^e u_{n,i_2}^{e\ast} f'(E_n-E) \right. \nonumber \\
 & & \left. + v_{n,i_1}^e v_{n,i_2}^{e\ast} f'(E_n + E) \right\}. \label{ldos_average_rewritten}
\end{eqnarray}

Fig.~\ref{ldos_temperature} shows the spatial-averaged LDOS around the Fermi energy [$|E/\Delta_0| \leq 4.0$] at $T/T_c = 0.1,~0.2,~0.3,~0.5,$ and $1.0$.
At $T/T_c = 0.1$, we can find some sharp peaks in the LDOS.
As the previous discussion, these peaks come from the discretization of energy levels.
In our approach, we obtain both positive and negative energy levels when we solve the BdG equations [Eqs.~(\ref{bdg_fem_1}) and (\ref{bdg_fem_2})].
Energy levels are symmetric with respect to $E=0$, but magnitudes of the LDOS are asymmetric.
This asymmetric magnitude of the LDOS results from the fact that $E = 0$ corresponds to the Fermi energy in the system.
In our calculation in Eqs.~(\ref{bdg_fem_1}) and (\ref{bdg_fem_2}), energy levels over the Fermi energy are also obtained.
Electric states below a bottom of energy levels in the negative energy region cannot exist.
So, the difference of the LDOS between the positive and the negative energy regions occurs.

Then, we consider behaviors of the LDOS with increasing the temperature.
At $T/T_c = 0.2,~0.3$, and $0.5$, peaks in the LDOS at $T/T_c = 0.1$ decrease or disappear.
Locations of peaks are almost same as peaks at $T/T_c = 0.1$, which means that values of energy levels do not depend on the temperature except for the very low temperature and the high temperature.
Finally, there are no peaks in the LDOS at $T/T_c = 1.0$.
Influences of the discretization of energy levels and the temperature are included in the derivative of the Fermi function $f'(E)$ in Eqs.~(\ref{ldos}) or (\ref{ldos_average_rewritten}).
Mathematically, this function behaves as a delta function in the low temperature, so sharp peaks appear in the LDOS.
Increasing the temperature, a width of the peak in the LDOS is spread.
Physically, the thermal energy becomes large when the temperature increases, and the influence of one energy level is spread to the large energy range.
When each peak is spread, neighbor peaks are superposed each other and small peaks disappear. 
Therefore, magnitudes of peaks decrease and the number of peaks becomes small with increasing the temperature.
This result means that the influence of the discretization of energy levels becomes small at the high temperature.
The suppression of the discretization comes from the fact that a thermal fluctuation suppresses the quantum confinement effect at the high temperature \cite{halperin}.

\subsection{\label{ldos_size_sec} Effects of size and shape on the LDOS}
\begin{figure}[tb]
\centering
\includegraphics[scale=0.49]{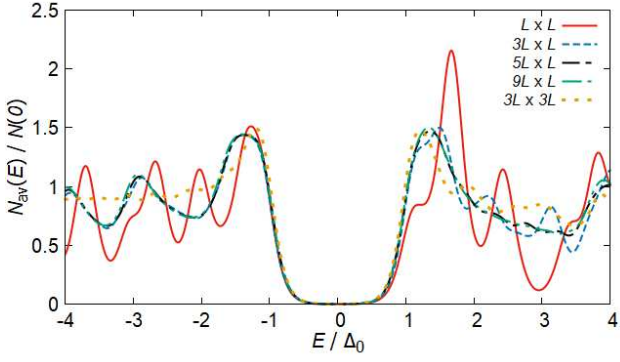}
\caption{\label{ldos_size} Spatial-averaged LDOS as a function of the energy in $L \times L$ (red), $3L \times L$ (blue), $5L \times L$ (black), $9L \times L$ (green), and $3L \times 3L$ (yellow) in $|E/\Delta_0| \leq 4.0$ at $T/T_c=0.2$.
LDOS plots in $3L \times L$, $5L \times L$, and $9L \times L$ are almost overlapped each other in the negative energy region ($E/\Delta_0 < 0.0$). 
In the positive energy region ($0.0 < E/\Delta_0 < 4.0$), LDOS plots in $5L \times L$ and $9L \times L$ are also almost overlapped each other.}
\end{figure}
\begin{figure}[htb]
\begin{center}
\includegraphics[scale=0.43]{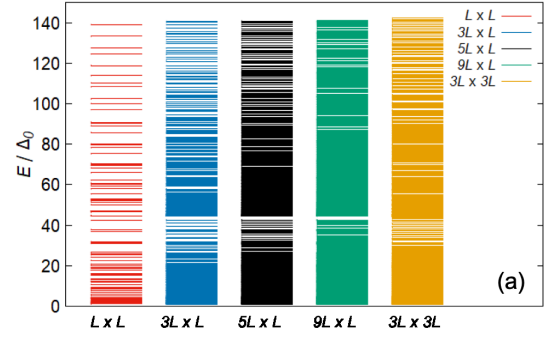}
\label{energy_level_a}
\end{center}
\begin{center}
\includegraphics[scale=0.43]{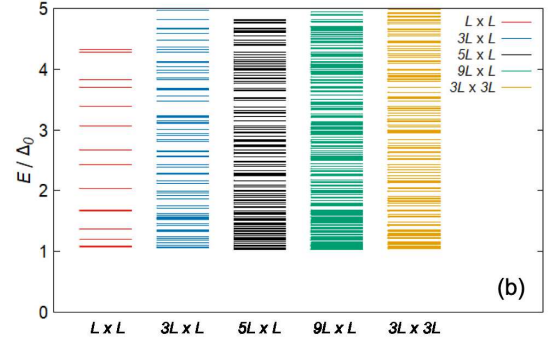}
\label{energy_level_b}
\end{center}
\begin{center}
\includegraphics[scale=0.43]{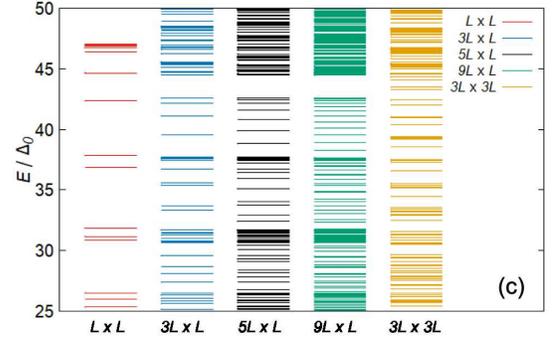}
\label{energy_level_c}
\end{center}
\caption{\label{energy_level} Energy levels in systems whose sizes are $L \times L$ (red), $3L \times L$ (blue), $5L \times L$ (black), $9L \times L$ (green), and $3L \times 3L$ (yellow) at $T/T_c = 0.2$. Energy levels are normalized to $\Delta_0$. Each plot is classified by the energy regions (a) $0 < E/\Delta_0 < 140.0$ [where all positive energy levels are included], (b) $0.0 < E/\Delta_0 < 5.0$ [where positive energy levels below the cutoff energy $E_c$ are included], and (c) $25.0 < E/\Delta_0 < 50.0$ [energy region where a maximum interval between neighbor energy levels in rectangular systems is included], respectively. A maximum value of the longitudinal axis in plot (b) corresponds to the cutoff energy, $E/E_c = (E/\Delta_0)\times(\Delta_0/E_c) = 1.0$}
\end{figure}

\begin{figure}[tb]
\centering
\includegraphics[scale=0.49]{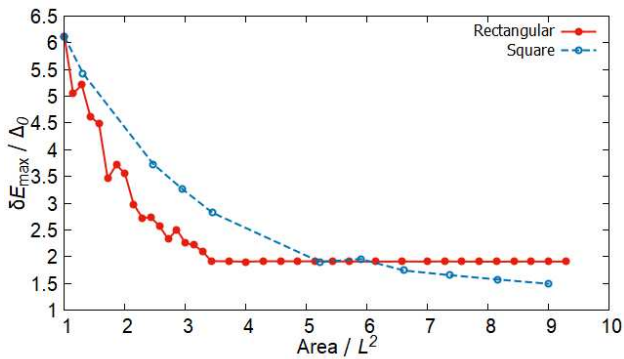}
\caption{\label{interval} Maximum interval between neighbor energy levels $\delta E_{\rm max}$ in the rectangular (red and solid line) and the square (blue and dashed line) systems at $T/T_c = 0.2$. The horizontal axis represents the area in the system divided by $L^2$. Intervals are normalized to $\Delta_0$.}
\end{figure}

\begin{figure*}[tb]
\begin{tabular}{cc}
\begin{minipage}{0.45\hsize}
\begin{center}
\includegraphics[width=\hsize]{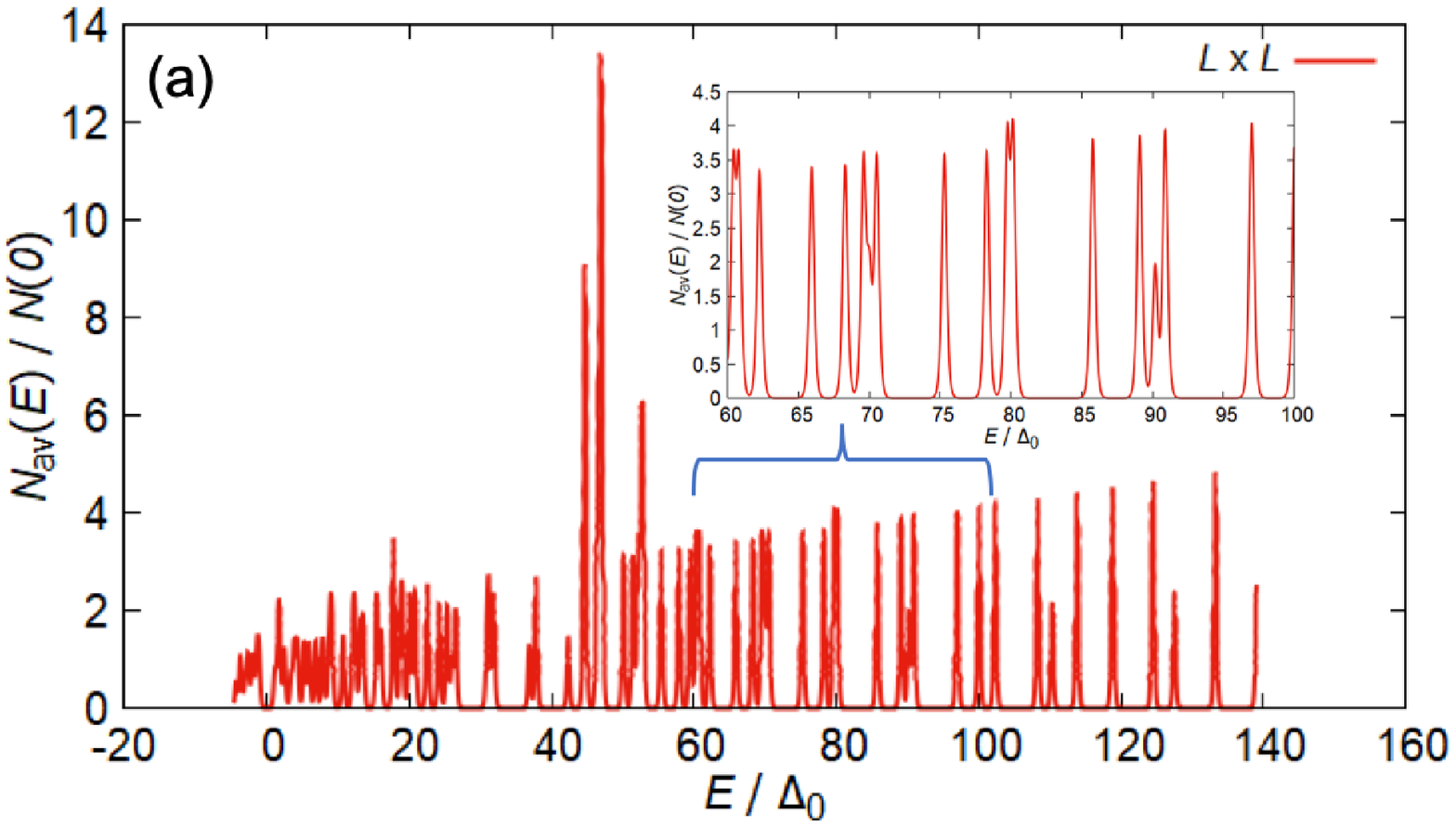}
\label{ldos_all_a}
\end{center}
\end{minipage}
\begin{minipage}{0.45\hsize}
\begin{center}
\includegraphics[width=\hsize]{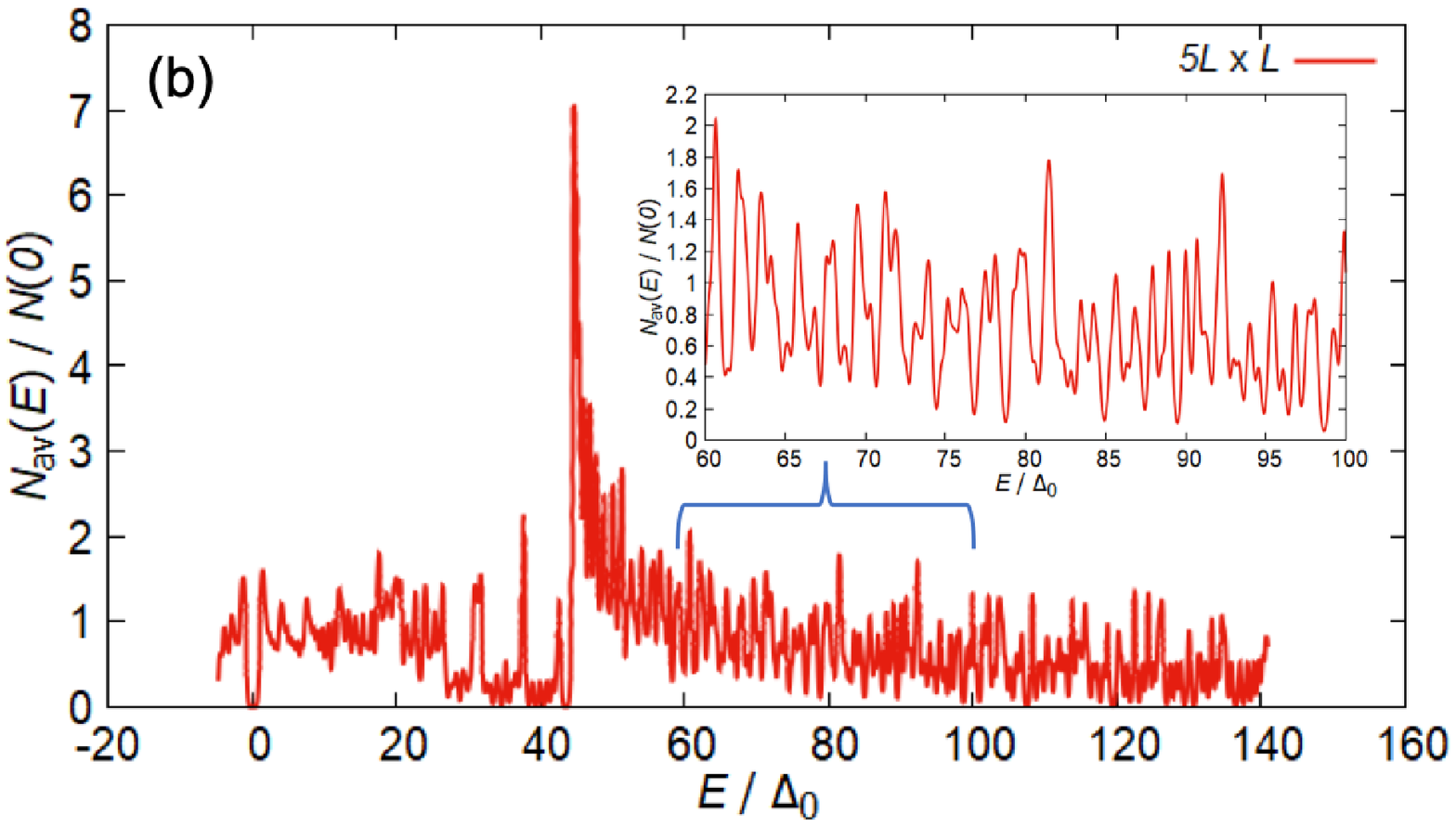}
\label{ldos_alll_b}
\end{center}
\end{minipage}
\end{tabular}
\begin{tabular}{cc}
\begin{minipage}{0.45\hsize}
\begin{center}
\includegraphics[width=\hsize]{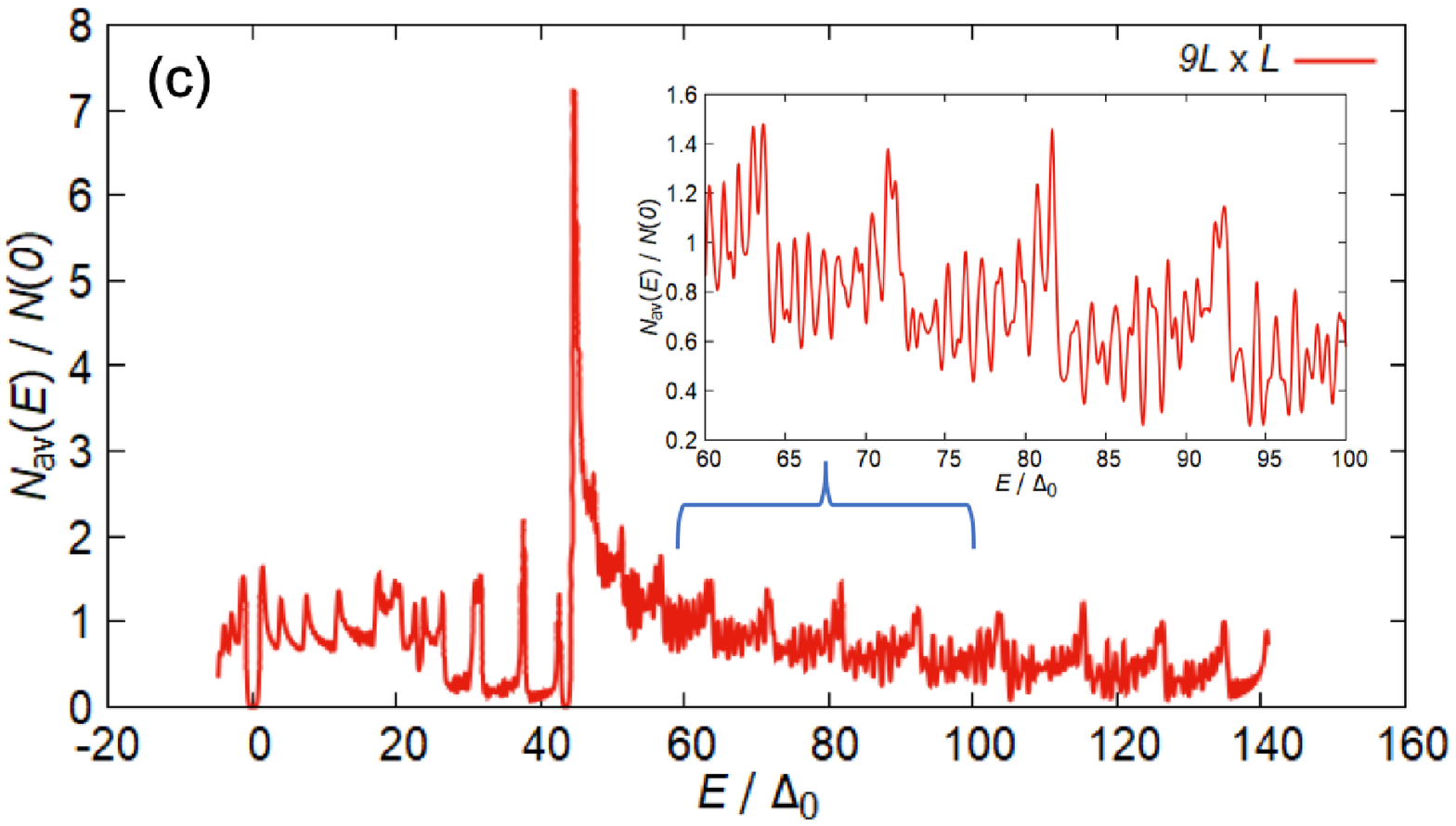}
\label{ldos_all_c}
\end{center}
\end{minipage}
\begin{minipage}{0.45\hsize}
\begin{center}
\includegraphics[width=\hsize]{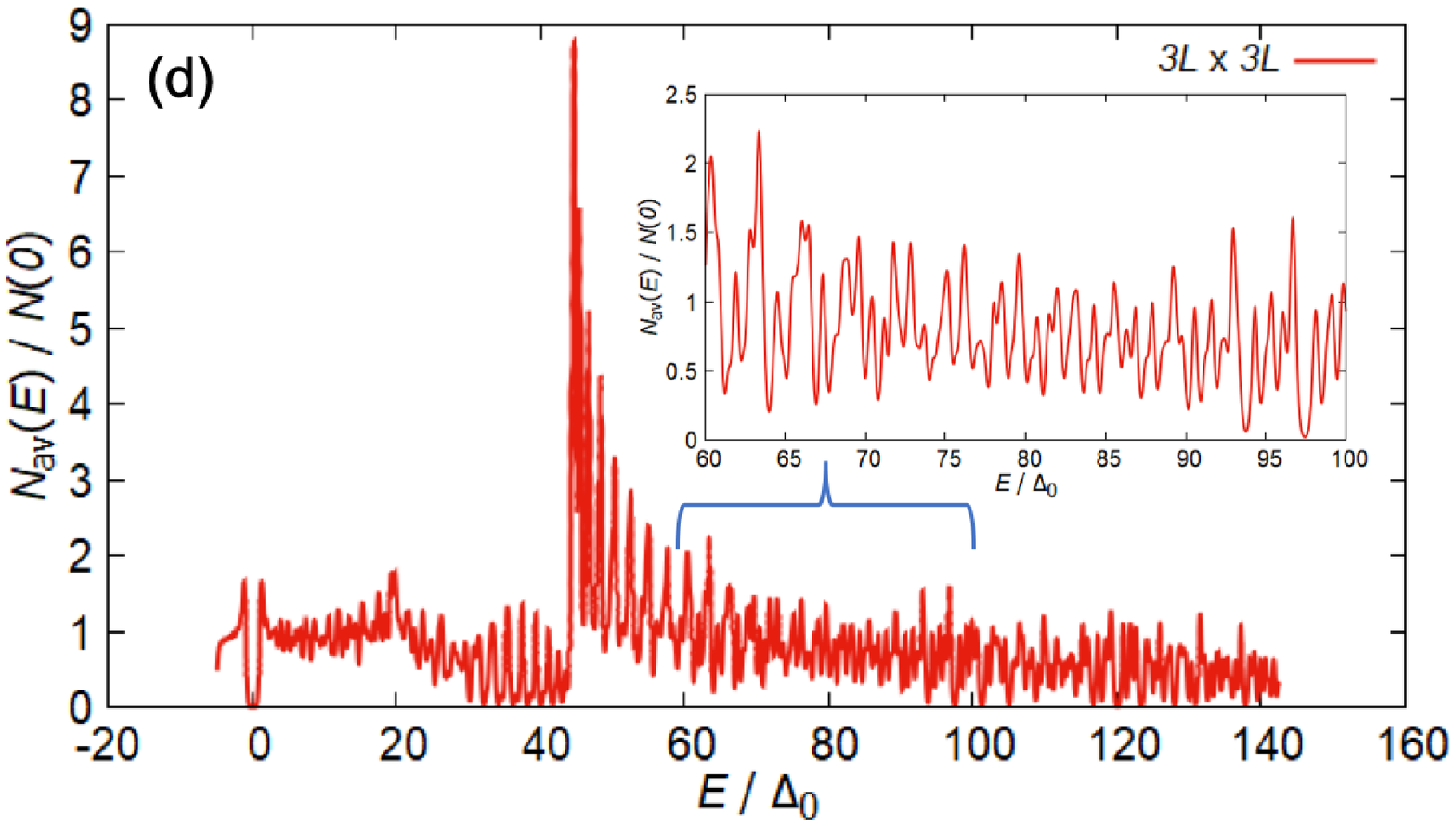}
\label{ldos_all_d}
\end{center}
\end{minipage}
\end{tabular}
\caption{\label{ldos_all} Spatial-averaged LDOS for all energy levels at $T/T_c=0.2$ in (a) $L \times L$, (b) $5L \times L$, (c) $9L \times L$, and (d) $3L \times 3L$ systems, respectively. Insert figures are LDOS plots in $60.0< E/\Delta_0 < 100.0$. }
\end{figure*}

In the previous subsection, the dependence of the temperature on the LDOS is investigated.
Next, we investigate dependences of the LDOS on the size and the shape in the system.
As well as the previous subsection, we calculate the spatial-averaged LDOS in Eq.~(\ref{ldos_average_rewritten}).
The temperature is fixed to $T/T_c = 0.2$ and the system size is changed.
The standard system size is $L \times L$ in Fig.~\ref{system_finite_element}(a) and we focus on rectangular systems and square systems.
In rectangular systems, a lateral length [$x$-axis side] is increased, but a longitudinal length [$y$-axis side] is fixed $L$.  
On the other hand, in square systems, both the lateral and the longitudinal length are increased equally.
 
Fig.~\ref{ldos_size} shows the LDOS around the Fermi energy [$|E/\Delta_0| \leq 4.0$] in $L \times L$, $3L \times L$, $5L \times L$, $9L \times L$, and $3L \times 3L$.
The LDOS in the $L \times L$ system is the same as the LDOS plot at $T/T_c = 0.2$ in Fig.~\ref{ldos_temperature}.
In the standard system size $L \times L$, there are many peaks in $|E/\Delta_0| < 4.0$ because of the discretization of energy levels as well as the previous discussion. 
At first, we concentrate on the LDOS in rectangular systems.
In comparison of the LDOS in the $L \times L$ with the $3L \times L$, the number of peaks in the $3L \times L$ system is less than the case in the $L \times L$ system.
Also, magnitudes of peaks decrease with increasing the lateral length.
Next, we compare the LDOS in the $3L \times L$ with the $5L \times L$.
Peaks in the negative energy region [$E / \Delta_0 < 0.0$] are not changed.
In the positive energy region [$E / \Delta_0 > 0.0$], magnitudes of peaks in the $5L \times L$ are smaller than the case in the $3L \times L$.
In both the $3L \times L$ and the $5L \times L$, a large peak of the LDOS in $E/\Delta_0 \sim 4.0$ appears. 
Finally, in the $9L \times L$ system, the LDOS in the negative energy region remains the same behavior as other rectangular cases [$3L \times L$ and $5L \times L$ systems].
On the other hand, peaks of the LDOS in the positive energy region $0 < E/\Delta_0 < 4.0$ almost disappear and a large peak appears in $E/\Delta_0 \sim 4.0$ as well as cases in the $3L \times L$ and the $5L \times L$ systems.  
Therefore, increasing only the lateral length, the number of peaks decreases but peaks do not disappear completely. 

Next, we compare the LDOS in the rectangular system with that in the square system.
The $9L \times L$ system and the $3L \times 3L$ system have same areas $9L^2$ and different shapes.
In the $3L \times 3L$ square system, there are no peaks of the LDOS in the negative energy region.
In the positive energy region $0 < E/\Delta_0 < 4.0$, peaks of the LDOS in the $3L \times 3L$ system are larger than peaks in the $9L \times L$ system.
However, a peak of the LDOS in $E/\Delta_0 \sim 4.0$ in the $3L \times 3L$ system is smaller than the peak of the LDOS in the same energy in the $9L \times L$ system. 

Clearly, the LDOS in the square system is different from the LDOS in the rectangular system with the same area.
We relate the difference of the LDOS to the quantum confinement effect.
The quantum confinement effect occurs when electrons are confined in the nano-scaled or atomic-scaled system.
When the system size increases, this confinement is suppressed.
In the case of the rectangular system, increasing the lateral length [$x$-axis side], the confinement of electrons from $x$-axis side is suppressed.
However, the longitudinal length [$y$-axis side] is fixed, then the confinement from $y$-axis side remains strong.
So, in the rectangular system, some large peaks still exist because of the strong confinement of electrons from the longitudinal side.
On the other hand, in the case of the square system, contributions of the confinement of electrons from both sides are suppressed equally.
Then, peaks of the LDOS in $E/\Delta_0 < 0.0$, which do not disappear in the case of the rectangular system, disappear because the confinement from the longitudinal side becomes small.
In $0 < E/\Delta_0 < 4.0$, small peaks of the LDOS still exist because the contribution of the confinement from the lateral side is larger than the contribution in the case of the rectangular system with the same area.
On the contrary, the small peak of the LDOS in $E/\Delta_0 \sim 4.0$ in the square system comes from the suppression of the confinement of electrons from the longitudinal side.
Therefore, the nano-scaled rectangular system with the constant longitudinal side will remain the quantum confinement effect unless the longitudinal length increases, while the quantum confinement effect in the nano-scaled square system will disappear with increasing lengths of both sides finally.

In Sec.~\ref{op_ldos}, we mentioned that the LDOS in the nano-scaled system results from the discretization of energy levels.
Next, we discuss the relation between system conditions and the discretization of energy levels in detail.
Fig.~\ref{energy_level} shows positive energy levels in $L \times L$, $3L \times L$, $5L \times L$, $9L \times L$, and $3L \times 3L$ at $T = 0.2T_c$.
Fig.~\ref{energy_level}(a) provides all positive energy levels in $0.0 < E /\Delta_0 < 140.0$, and Figs.~\ref{energy_level}(b) and (c) provide energy levels within certain energy regions; (b) $0.0 < E/\Delta_0 < 5.0$, and (c) $25.0 < E/\Delta_0 < 50.0$, respectively.
The energy level is normalized to $\Delta_0$.
The maximum value of the longitudinal axis in Fig.~\ref{energy_level}(b) corresponds to the cutoff energy, $E/E_c = (E/\Delta_0) \times (\Delta_0/E_c) = 1.0$.
Maximum and minimum values of energy levels in the positive energy region are almost same, but slightly change with increasing the system size.
It is known that this small change of energy levels occurs in nano-scaled superconductors due to the quantum confinement effect \cite{umeda}.
In the standard system $L \times L$, intervals between neighbor energy levels are large.
When the system size increases, new energy levels enter in the energy region $0.0 < E/\Delta_0 < 140.0$ and do not enter above the maximum value of the energy level and below the gap energy.
In the case of rectangular systems, a large interval between neighbor energy levels appears in the energy region $40.0 < E/\Delta_0 < 45.0$ [See Fig.~\ref{energy_level}(b)].
Even in the $9L \times L$ system, energy levels do not exist within this large interval.
On the other hand, in the case of the square system $3L \times 3L$, some energy levels exist in $40.0 < E/\Delta_0 < 45.0$.
So, the interval between neighbor energy levels in this region is shorter than the interval in the rectangular system.
Dependences of the maximum interval between neighbor energy levels $\delta E_{\rm max}$ on the size and the shape in the rectangular and the square systems are shown in Fig.~\ref{interval}.
The horizontal axis in Fig.~\ref{interval} provides each area divided the area in the standard system size $L \times L$.
In the rectangular case in Fig.~\ref{interval}, the maximum interval becomes almost constant when the area increases.
Constant intervals in large areas correspond to maximum intervals in the energy region  $40.0 < E/\Delta_0 < 45.0$ in Fig.~\ref{energy_level}(c).
On the contrary, the maximum interval in the square system continues to decrease even when the area is large.
This result means that energy levels in the square system can be arranged more freely, and the energy interval in $40.0 < E/ \Delta_0 < 45.0$ is no longer the maximum interval in the square system.

Below the largest interval between neighbor energy levels in rectangular system in $40.0 < E/\Delta_0 < 45.0$, we find periodic arrangements of energy levels [See Fig,~\ref{energy_level}(c)].
In rectangular systems, big packets of energy levels and some discrete energy levels appear periodically in $25.0 < E/\Delta_0 < 50.0$.
As the previous discussion of the confinement of electrons, increasing the lateral length, the confinement of electrons from the lateral side is suppressed, while the confinement from the longitudinal side remains strong.
Contributions of the confinement are different between the lateral and the longitudinal side, which leads to the periodic arrangement of big packets and some discrete energy levels in rectangular systems.
On the contrary, in the square system, packets of energy levels are smaller than packets in the rectangular system, while there is an arrangement of energy levels with almost equal intervals in $30.0 < E/\Delta_0 < 45.0$.
Equally suppressions of the confinement of electrons from both sides result in this regular arrangement.
In comparison of two different energy regions in Figs.~\ref{energy_level}(b) and (c), a density of energy levels below the cutoff energy in Fig.~\ref{energy_level}(b) is large.
Below the cutoff energy, big packets are created in rectangular systems, while in the square system, small packets are created and intervals between neighbor energy levels are relatively equal as well as the arrangement in $25.0 < E/\Delta_0 < 50.0$.
Therefore, a tendency of arrangements of energy levels in different energy regions is the same.

So far, we restricted our discussion of the LDOS to the energy region around the Fermi energy [$|E/\Delta_0| < 4.0$, in Figs.~\ref{ldos_temperature} and \ref{ldos_size}]. 
Finally, we extend our discussion to the energy region included all energy levels.
Fig.~\ref{ldos_all} shows LDOS plots within all energy levels in (a) $L \times L$, (b) $5L \times L$, (c) $9L \times L$, and (d) $3L \times 3L$.
The horizontal axis represents the energy normalized to $\Delta_0$.
LDOS plots in $60 < E/\Delta_0 < 100$ are shown in insert figures in order to emphasize periodic behaviors of the LDOS.
Also, the left edge of each plot corresponds to the bottom of energy levels.
In all figures, a very large peak appears in $E/\Delta_0 \sim 45$.
This value corresponds to the energy above the maximum interval between energy levels in rectangular systems [see Fig.~\ref{energy_level}(c)].

We discuss dependences of peak structures of the LDOS in all energy levels on the size and the shape.
In Fig.~\ref{ldos_all}(a), the LDOS has many sharp peaks locally.
From Fig.~\ref{energy_level}(a), intervals between neighbor energy levels are large, and superpositions of neighbor peaks in the LDOS hardly occur.
In rectangular systems [Figs.~\ref{ldos_all}(b) and (c)], some ensembles with periodic behaviors appear in $0 < E/\Delta_0 < 20.0$ and $50.0 < E/\Delta_0 < 140.0$.
In the periodic arrangement of energy levels in $0 < E/\Delta_0 < 20.0$, the LDOS has a locally largest peak in $E/\Delta_0 \sim 4.0,~7.4,~11.8$, and $17.5$ except for the energy corresponding to the gap energy.
Then, the LDOS decreases like a function of the minus square root with increasing the energy.
On the other hand, in $50.0 < E / \Delta_0 < 140$, locally largest peaks in the LDOS appear in $E/\Delta_0 \sim 63.6,~71.4,~81.6,~92.4$, and so on as well as the case in $0 < E / \Delta_0 < 20.0$ [see the insert figure in Fig.~\ref{ldos_all}(c)].
However, the LDOS oscillates and increases like the function of the square with increasing the energy, then the LDOS drops in next peaks.
It is known that the LDOS has a multi gap structure in the region of the quantum confinement effect \cite{chen,flammia}.
Therefore, these periodic peaks can be regarded as gaps in the multi gap structure due to the confinement of electrons from the fixed longitudinal side.
It is expected that these gaps will not disappear unless the longitudinal length increases.

From the LDOS of the square system in Fig.~\ref{ldos_all}(d), more peaks appear than the case in rectangular systems in Figs.~\ref{ldos_all}(b) and (c). 
In $50.0 < E / \Delta_0 < 140$, a small periodic structure in the LDOS appear [see the insert figure in Fig.~\ref{ldos_all}(d)].
As the previous discussion of the discretization of energy levels, confinements of electrons from both longitudinal and lateral sides are suppressed equally.
Repeatedly, in comparison of cases in the rectangular and the square systems with the same area [$9L \times L$ and $3L \times 3L$], the confinement of electrons from the lateral side in the $3L \times 3L$ system is larger than the confinement from the lateral side in the $9L \times L$ system.
Then, the peak structure with the small period is left in the LDOS.
It is expected that if the size of the square system increases, small peaks will disappear because of the superposition between neighbor peaks and only large peaks in $E/\Delta_0 \sim 1.0,~20.0,$ and $45.0$ will be left.
Finally, these large peaks in $E/\Delta_0 \sim 1.0,~20.0,$ and $45.0$ will disappear in the bulk limit.

\section{\label{summary} Summary}
We have investigated the local density of state (LDOS) in the two-dimensional nano-structured superconductor.
We have solved the Bogoliubov-de Gennes equations with the finite element method.
In the finite nano-structured system, energy levels are discrete, which lead to the discretization of the LDOS with respect to the energy.
This discretization of the LDOS depends on the temperature.
At the low temperature, the LDOS has many sharp peaks.
When the temperature increases, the influence of each energy level is spread to the large energy range and magnitudes of peaks in the LDOS decrease. 
Also, the LDOS also depends on the size and the shape of the system for the fixed temperature.
The number of peaks in the LDOS decreases when the system size increases. 
This behavior is different between the rectangular system and the square system.
In this time, the rectangular system with the nano-size fixed longitudinal length and the variable lateral length is considered.
In the rectangular system, when the lateral length increases, the confinement of electrons from the lateral side is suppressed.
Energy levels become continuous gradually, but the maximum interval between energy levels is saturated because the confinement of electrons from the longitudinal side remains strong.
Then, the number of the peak in the LDOS decreases, while peaks in the LDOS cannot disappear completely.
These peaks can be regarded as gaps in the multi gap structure due to the quantum confinement effect.
On contrary, in the square system, confinements of electrons from both lateral and longitudinal sides are suppressed equally.
The maximum interval between energy levels continues to decrease even in the large area.
Comparing with the LDOS in the rectangular system, more peaks appear in the LDOS in the square system.
Then, the LDOS has the peak structure with the small period. 
Finally, the peak structure of the LDOS in the square system will disappear in the bulk limit.
Therefore, differences of the shape and size affect behaviors of the LDOS and energy levels.

In this research, we focus on the rectangular and the square systems.
We do not take into account the thickness of the system.
As Sec.~\ref{introduction}, it is known that the change of the thickness of the film in the nano-structured superconductor occurs the oscillation of superconducting properties.
If we consider cases that influences of the thickness can be neglected [for example, systems with nano-sized fixed thickness or atomic-layer thickness], the same discussion will be valid.
Of course, exactly, contributions from the direction of thickness is cannot be neglected.
In order to consider influences of the thickness on the LDOS, we have to investigate the LDOS in three-dimensional superconductors.
Also, we have considered the conventional superconducting plane.
Different behaviors of the LDOS are expected in cases of junction systems with non-superconducting samples and other shapes such as a triangle, a circle, and so on.
In particular, in junction systems such as superconductor-normal metal-superconductor (SNS) junction systems \cite{hsiang,blaauboer,chtchelkatchev} and superconductor-ferromagnet-superconductor (SFS) junction systems \cite{buzdin,halterman_prb_2015,halterman_sust_2016}, the LDOS is more complicated because of the Andreev reflection and the proximity effect.
The nano-structured Josephson junction system is expected as applications such as the nano-SQUID, the quantum computer, and the single flux quantum (SFQ) logic.
In the future work, dependences of the LDOS and other properties in nano-structured SNS junctions or SFS junctions on the discretization of energy levels due to the quantum confinement effect will be investigated.


\bibliography{reference.bib}

\end{document}